\documentclass[12pt,a4paper,twoside]{article}
\RequirePackage{latexsym,amsmath,amssymb}
\RequirePackage[dvipsnames,usenames]{color}
\usepackage{a4wide}
\usepackage{amssymb}
\usepackage{bm}
\usepackage{dcolumn}
\usepackage{amsfonts}
\usepackage{graphicx,epsfig}
\usepackage{psfrag}
\usepackage{cite}
\usepackage{multicol}
\usepackage{tikz}
\usetikzlibrary{arrows,shapes}
\usetikzlibrary{matrix,arrows}
\newcommand{\beq}{\begin{eqnarray}}
\newcommand{\eeq}{\end{eqnarray}}
\newcommand{\be}{\begin{equation}}
\newcommand{\ee}{\end{equation}}

\newcommand{\ket}[1]{\mbox{$\mid #1\,\rangle$}}

\newcommand{\expec}[1]{\mbox{$\langle\, #1\,\rangle$}}

\renewcommand{\d}{\mbox{${\rm d}$}} 
\newcommand{\lp}{\ell_{\rm p}}
\newcommand{\mpl}{m_{\rm p}}
\newcommand{\gn}{G_{\rm N}}
\newcommand{\rh}{r_{\rm H}}

\newcommand{\Rh}{R_{\rm H}}

\title{\bf Horizon Quantum Mechanics for spheroidal sources}
\author{Roberto~Casadio$^{ab}$\thanks{E-mail: casadio@bo.infn.it}
$\ ,$
Andrea~Giusti$^{abc}$\thanks{E-mail: andrea.giusti@bo.infn.it}
$\ $
and
Rehana~Rahim$^{ad}$\thanks{rehana.rahim8@gmail.com}
\\
\\
$^a${\em Dipartimento di Fisica e Astronomia, Universit\`a di Bologna}
\\
{\em via Irnerio~46, I-40126 Bologna, Italy}
\\
\\
$^b${\em I.N.F.N., Sezione di Bologna, IS - FLAG}
\\
{\em via B.~Pichat~6/2, I-40127 Bologna, Italy}
\\
\\
$^c${\em Arnold Sommerfeld Center, Ludwig-Maximilians-Universit\"at}
\\
{\em Theresienstra{\ss}e 37, 80333 M\"unchen, Germany}
\\
\\
$^d${\em Department of Mathematics, Quaid-i-Azam University}
\\
{\em Islamabad, Pakistan}
}
\begin{document}
\maketitle
\begin{abstract}
We start investigating the extension of the Horizon Quantum Mechanics to the case of non-spherical
sources.
We first study the location of trapping surfaces in space-times resulting from an axial deformation
of static isotropic systems, and show that the Misner-Sharp mass evaluated on the corresponding
undeformed spherically symmetric space provides the correct gravitational radius to locate
the horizon.
We finally propose a way to determine the deformation parameter in the quantum theory.
\end{abstract}
\section{Introduction and motivation}
\label{intro}
\setcounter{equation}{0}
In General Relativity, black holes are portions of Lorentzian manifolds characterised
by the existence of an event horizon~\cite{Wald,HE}.
The prototype is of course given by the Schwarzschild metric~\cite{schwarz}, 
\be
\d s^2
=
-\left(1-\frac{\Rh}{r}\right) \d t^2
+\left(1-\frac{\Rh}{r}\right)^{-1} \d r^2
+
r^2
\left(\d\theta^2+\sin^2\theta\,\d\phi^2\right)
\ ,
\label{schw}
\ee
which was discovered right after Einstein's formulation of his gravity theory and is fully characterised
by the gravitational radius $\Rh=2\,\gn\,M$.
In more general gravitating systems, the local counterpart of the event horizon is given by
a trapping surface, which can be naively understood as the location where the escape
velocity equals the speed of light at a given instant.
If the system is spherically symmetric, one can also introduce a quasi-local mass function
$m=m(t,r)$ and a local gravitational radius $\rh(t,r)=2\,\gn\,m(t,r)$, from which the location of trapping
surfaces can be determined~\cite{ashtekar}.
\par
The Hawking radiation~\cite{hawking} was later discovered in the semiclassical picture of
matter fields quantised on the manifold~\eqref{schw}, and has raised paradoxes indicating
a possible incompatibility between the quantum theory and General Relativity. 
In this respect, it appears useful to remark that quantities like the Arnowitt-Deser-Misner (ADM)
mass $M$ and the Misner-Sharp mass function $m$ effectively encode the crucial physics
of black holes.
In quantum physics, the energy density that defines the Misner-Sharp mass $m$
(and ADM mass $M$) becomes a quantum observable and one could conjecture the gravitational radius
admits a similar description. 
The horizon quantum mechanics (HQM) was in fact proposed~\cite{fuzzyh} in order to describe the
``fuzzy'' Schwarzschild (or gravitational) radius of a localised, isotropic and static, quantum source.
Unlike most other attempts, in which the gravitational degrees of freedom are quantised independently,
this approach lifts the relation between the state of the source and the state of the gravitational radius
to a (local or global) quantum constraint~\cite{hqm, Giusti:2017byx}.
The HQM has already been applied to several cases which can be reduced to isotropic
sources~\cite{gupf,Tevo,cx,otherD,acmo,cc, Giugno:2017xtl}.
However, its extension to non-spherical systems requires one to identify a mass
function from which the location of trapping surfaces can be uniquely determined
and which depends only on the state of the matter source, like the Misner-Sharp
mass for isotropic sources.
The latter property is crucial in a perspective in which any geometric properties
must be eventually recovered from the quantum state of the whole matter-gravity
system.
For instance, the global HQM for rotating sources was introduced in Ref.~\cite{rotating} by taking
advantage of the asymptotic Kerr charges.
\par
In this paper, we shall consider how the construction can be carried out when the system
has slightly spheroidal symmetry.
For this purpose, we shall deform a spherically symmetric space-time and study the 
location of trapping surfaces perturbatively in the deformation parameter.
Finally, we outline how the deformation can be characterised quantum mechanically.
\section{Spheroidal sources}
\label{Sec:spheroidal}
\setcounter{equation}{0}
The study of trapping surfaces is a very complex topic, and determining their existence and
location is in general very difficult. 
In this section we will investigate how the description of trapping surfaces for static spherically
symmetric systems can be extended to the case in which the symmetry is associated with (slightly)
spheroidal surfaces.
\par
A static and spherically symmetric metric $g_{\mu\nu}$ (with $\mu,\nu=0,\ldots,3$) can always
be written as
\be
\d s^2
=
-A(r)\,\d t^2+\frac{\d r^2}{B(r)}
+
r^2
\left(\d\theta^2+\sin^2\theta\,\d\phi^2\right)
\ ,
\label{metric}
\ee
where $r$ is the areal coordinate of the spheres parameterised by $\theta$ and $\phi$.
A trapping surface then exists where the expansion of outgoing null geodesics vanishes.
In general, the expansion scalars associated with outgoing and ingoing geodesics
are respectively given by
\begin{equation}
\Theta_{\bm{\ell}}
=
q^{\mu\nu}\,\nabla_\mu l_{\nu}
\ ,
\qquad
\Theta_{\bm{n}}
=q
^{\mu\nu}\,\nabla_\mu n_{\nu}
\ ,
\label{h10}
\end{equation}
where $q_{\mu \nu}=g_{\mu \nu}+l_{\mu}\,n_{\nu}+n_{\mu}\,l_{\nu}$
is the metric induced by $g_{\mu\nu}$ on the two-dimensional space-like
surface formed by spatial foliations of the null hypersurface generated by
the outgoing tangent vector $\bm{\ell}$ and the ingoing tangent vector $\bm{n}$.
This 2-dimensional metric is purely spatial and has the following properties
\begin{equation}
q_{\mu\nu}\,\ell^{\mu}
=
q_{\mu\nu}\,\ell^{\nu}=0
\ ,
\quad
q^\mu_{\ \mu}=2
\ ,
\quad
q^\mu_{\ \lambda}\,q^\lambda_{\ \nu}
=
q^\mu_{\ \nu}
\ ,
\label{h11}
\end{equation}
where $q^\mu_{\ \nu}$ represents the projection operator onto the 2-space orthogonal
to $\bm{\ell}$.
For the metric~\eqref{metric}, one then finds that the product
\be
\Theta_{\bm{\ell}} \,\Theta_{\bm{n}}
\propto
g^{ij}\,\nabla_i r\,\nabla_j r
\ee
vanishes on trapped surfaces. 
\par
If $\rho=\rho(r)$ is the isotropic energy density of the source~\footnote{We shall always assume that
the matter source also contains a (isotropic) pressure term such that the Tolman-Oppenheimer-Volkov
equation of hydrostatic equilibrium is satisfied~\cite{stephani}.},
Einstein's field equations yield
\be
B(r)=1-\frac{\rh(r)}{r}
\ ,
\label{grr}
\ee
where~\footnote{We shall use units with $\gn=c=1$ in this and the next section.}
\be
\rh(r)
=
2\,{m(r)}
\label{hoop}
\ee
is the gravitational radius determined by the Misner-Sharp mass function
\be
m(r)
=
4\,\pi
\int_0^r \rho(\bar r)\,\bar r^2\,\d \bar r
\ .
\label{M}
\ee
A horizon then exists where $B=0$, or where the gravitational radius
satisfies 
\be
\rh(r)= r
\ ,
\label{Ehor}
\ee
for $r>0$.
If the source is surrounded by the vacuum, the Misner-Sharp mass asymptotically approaches the
Arnowitt-Deser-Misner (ADM) mass of the source, $m(r\to\infty)=M$,
and the gravitational radius likewise becomes the Schwarzschild radius $\Rh= 2\,M$.
To summarise, the relevant properties of the Misner-Sharp mass~\eqref{M} are that it only
depends on the source energy density and allows one to locate the trapping surfaces via
Eq.~\eqref{hoop}.
\par
\begin{figure}[t!]	
\centering
\includegraphics[scale=0.5]{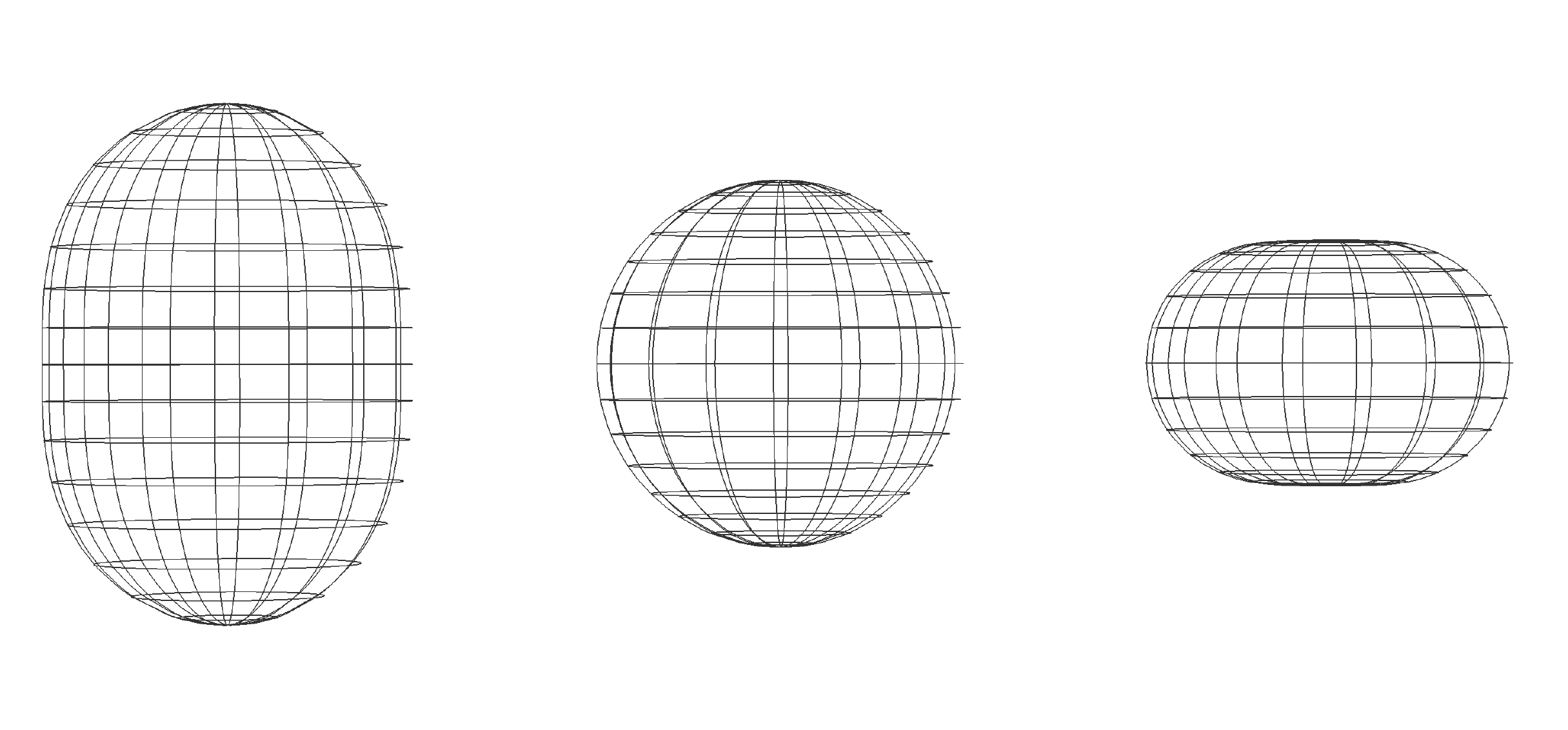}
\caption{Spheroids: prolate spheroid with $a^2>0$ (left) compared to oblate spheroid with $a^2<0$
(right) and to the reference sphere $a^2=0$ (centre).}
\label{f1}
\end{figure}

We now change to (oblate or prolate) spheroidal coordinates and consider a localised source
of spheroidal radius $r=r_0$, say with mass $M_0$, surrounded by a fluid with energy
density $\rho=\rho(r)$.
The central source only serves the purpose to avoid discussing coordinate singularities
at $r=0$.
In the interesting portion of space $r>r_0$, we assume the metric $g_{\mu\nu}$ is of the form
\begin{equation}
\d s^{2}
=
-h_a(r,x)\, \d t^{2}
+
\frac{r^{2}+a^{2}\,x^{2}}{r^{2}+a^{2}}\,
\frac{\d r^{2}}{h_a(r,x)} 
+\frac{r^{2}+a^{2}\,x^{2}}{1-x^2}\,\d x^{2}
+\left(r^{2}+a^{2}\right)\left(1-x^2\right) \d\phi^{2}
\ ,
\label{1}
\end{equation}
where $h_a=h_a(r,x\equiv\cos\theta)$ is a function to be determined.
Surfaces of constant $r$ now represent ellipsoids of revolution, or spheroids, on which the
density is constant.
For $a^2>0$, the above metric can describe the space-time outside a prolate spheroidal source,
which extends more along the axis of symmetry than the equatorial plane (see left surface in Fig.~\ref{f1}).
In order to describe an oblate source, which is flatter along the axis (see right surface in Fig.~\ref{f1}),
we can simply consider the mapping $a\to i\,a$ (so that $a^2\to -a^2$).
\par
The energy-momentum tensor $T_{\mu\nu}$ of the source can be inferred from the Einstein equations,
\begin{equation}
G_{\mu\nu}
=
R_{\mu\nu}
-\frac{1}{2}\,R\,g_{\mu\nu}
=8\,\pi\,T_{\mu\nu}
\ ,
\label{2}
\end{equation}
where $G_{\mu\nu}$ is the Einstein tensor, $R_{\mu\nu}$ the Ricci tensor and
$R$ the Ricci scalar.
However, we are only interested in ensuring that the energy density is spheroidally symmetric,
that is
\be
G^0_{\ 0}
=
-8\,\pi\,\rho(r)
\ ,
\label{eG00}
\ee
and we will therefore assume the necessary pressure terms are present in order to 
maintain equilibrium.
Given the symmetry of the system, we can restrict the analysis to the upper half spatial volume 
$1\ge x\ge 0$ corresponding to $0\le\theta\le \pi/2$.
Since the Einstein tensor turns out to be rather involved,
we proceed by considering small departures from spherical symmetry, parameterised
by $a^2\ll r_0^2$, and expand all expressions up to order $a^2$.
In particular, the energy density must have the form
\be \label{rhoa-exp}
\rho
\simeq
\rho_{(0)}(r)
+
a^2\,\rho_{(2)}(r)
\ ,
\ee
whereas the unknown metric function
\beq \label{ha-exp}
h_a
&\!\!\simeq\!\!&
h_{(00)}(r)
+
a^2
\left[h_{(20)}(r)+x^2\,h_{(22)}(r)\right]
\ ,
\nonumber
\\
&\!\!\simeq\!\!&
1-\frac{2\,m_{(00)}}{r}
-
2\,a^2\,
\frac{m_{(20)}+x^2\,m_{(22)}}{r}
\ ,
\label{h}
\eeq
where we introduced a Misner-Sharp mass function $m_{(00)}=m_{(00)}(r)$,
like in Eq.~\eqref{M}, for the zero order term and corrective terms $m_{(2i)}=m_{(0i)}(r)$ at order $a^2$,
with $i = 0, 2$.
For the sake of clarity, the subscripts $(0)$ and $(2)$ represent the order of $a^2$
in Eq.~\eqref{rhoa-exp}, whereas with the subscripts $(ij)$ we denote a term of the expansion of $h_a$
that appears multiplied by $a^i \, x^j$ in Eq.~\eqref{ha-exp}.
\par	
At zero order, in fact, Eq.~\eqref{eG00} reads
\be
{G_{(0)}}^0_{\ 0}
=
-\frac{2\,m_{(00)}'}{r^2}
=
-8\,\pi\,\rho_{(0)}
\ ,
\label{m00}
\ee
with primes denoting derivatives with respect to $r$.
The solution $m_{(00)}$ is correctly given by the relation~\eqref{M}.
\par
At first order in $a^2$, the Einstein tensor contains two terms, 
\be
{G_{(2)}}^0_{\ 0}
=
F(r)
+x^2\,L(r)
\ ,
\ee
where $F (r)$ and $L(r)$ are independent of $x$.
Since $\rho$ does not depend on $x$ by construction, we must have $L= 0$, which yields
\be
m_{(22)}'
+
\left(1-\frac{2\,m_{(00)}}{r}\right)^{-1}\frac{3\,m_{(22)}}{r}
-
\frac{3}{2\,r^2}
\left(m_{(00)}'-\frac{5\,m_{(00)}}{3\,r}\right)
=
0
\ .
\label{m22}
\ee
Finally, we are left with
\be
F
=
-\frac{2\,m_{(20)}'}{r^2}
-
\frac{m_{(00)}'}{r^4}
+
\frac{3\,m_{(00)}}{r^5}
+
\frac{2\,m_{(22)}}{r^3}
\left(1-\frac{2\,m_{(00)}}{r}\right)^{-1}
=
-8\,\pi\,\rho_{(2)}
\ ,
\label{m20}
\ee
in which $m_{(00)}$ is determined by Eq.~\eqref{m00} and $m_{(22)}$ by Eq.~\eqref{m22}, respectively.
Eq.~\eqref{m20} can then be used to determine $m_{(20)}$.
\par
Once the metric function $h_a=h_a(r,x)$ is obtained, one can determine the locations of trapping surfaces 
from the vanishing of the expansion of outgoing null geodesics as defined in Eq.~\eqref{h10}.
It will then be interesting to compare the result with the solutions of the generalised Eq.~\eqref{hoop},
namely
\be
2\,m_a(\rh,x)=\rh(x)
\ ,
\label{MSrh}
\ee
where
\be
m_a(r,x)
\simeq
m_{(00)}(r)
+
a^2\left[m_{(20)}(r)+x^2\,m_{(22)}(r)\right]
\ ,
\qquad
\ee
is now the extended Misner-Sharp mass.
We also note that Eq.~\eqref{MSrh} is equivalent to
\be
h_a(\rh,x)
=
0
\ ,
\label{h=0}
\ee
which will be checked below with a specific example.
We can just anticipate that we expect the location of the trapping surface respect the
spheroidal symmetry of the system and is thus given by the spheroidal deformation
of the isotropic horizon obtained for $a\to 0$.
\section{A simple example: spheroidal de~Sitter}
\label{sec:3}
\setcounter{equation}{0}
We shall now apply the above general construction to a specific example, namely
\be
\rho(r)
=
\rho_{(0)}(r)
=
\frac{\alpha ^2}{4\,\pi\,r}
\ ,
\label{rhoEx1}
\ee
where $\alpha$ is a positive constant independent of $a$ (so that $\rho_{(2)}=0$).
%
%
%
\par
From Eq.~\eqref{m00}, we obtain
\be
m_{(00)}
=
M_0
+
\frac{\alpha^2 \, (r^2-r_0^2)}{2}
\ ,
\ee
which of course holds for $r>r_0$.
Since this case does not show any apparent spurious singularity at $r=0$, we further set
$\alpha^2\,r_0^2\simeq 2\,M_0$, so that
\be
m_{(00)}
\simeq
\frac{\alpha^2 \, r^2}{2}
\ .
\ee
At zero order in the deformation, we then find the usual horizon for the isotropic de~Sitter space
located at
\be
r_{\rm H}
=
2 \, m_{(00)}
=
\alpha^{-2}
\ .
\label{rha0}
\ee
\par
Next, Eq.~\eqref{m22} reads
\be
m_{(22)}'
+\frac{3\,m_{(22)}}{(1-\alpha ^2 \, r)\,r}
-\frac{\alpha^2}{4\,r}
=
0
\ ,
\ee
and admits the general solution
\be
m_{(22)}
=
\frac{1 - \alpha^2 \, r}{8\,\alpha^4\,r^3}
\left\{1 - \left[4 - 8\,\alpha^4\,\, (1 - \alpha^2 \, r)\,C_{(22)} 
+ 2 \, (1 - \alpha^2 \, r) \, \ln (1 - \alpha^2 \, r) \right] (1 - \alpha^2 \, r)
\right\}
\ ,
\ee
with $C_{(22)}$ an integration constant.
In particular, we notice that, for $r \simeq \alpha^{-2}$, the general solution reduces to
\be
m_{(22)}
\simeq
\frac{\alpha^2}{8} \, \left( 1 - \alpha^2 \, r \right)
\ .
\ee
We can then determine $m_{(20)}$ from Eq.~\eqref{m20}, which,
on employing the above expansion for $m_{(22)}$, reads
\be
m_{(20)}'
\simeq
\frac{3 \, \alpha^2}{8 \, r}
\ ,
\ee
and yields
\be
m_{(20)} (r)
\simeq
C_{(20)} + \frac{3 \, \alpha^2}{8} \, \ln( \alpha^2 r)
\ ,
\ee
with $C_{(20)}$ another integration constant.
\par
We then set $C_{(20)}=C_{(22)}=0$ for simplicity, and obtain
\be
m_a(r,x)
\simeq
\frac{\alpha^2 \, r^2}{2}
+ \frac{a^2 \, \alpha^2}{8} \, \left[ (1 - \alpha^2 \, r) \, x^2 + 3 \, \ln( \alpha^2 r) \right]
\ ,
\ee
again for $r\simeq \alpha^{-2}$.
After substituting $m_{(00)}$, $m_{(20)}$ and $m_{(22)}$ into Eq.~\eqref{h}, we have 
\be
h_a(r,x)
\simeq
\left(1-\alpha^{2}\,r\right)
-\frac{a^2\,\alpha^{2}}{4\,r}
\left[
3\,\log (\alpha^2 \, r)+x^2\,(1-\,\alpha^2\, r)
\right]
\ .
\quad
\label{h1}
\ee 
The condition~\eqref{h=0} then admits two solutions, namely the unperturbed
horizon in Eq.~\eqref{rha0} and
\be
r_{\rm H}
\simeq
\frac{a^2 \, \alpha ^2 \, (9 - 2 \, x^2)}{3\,a^2 \, \alpha^4 - 8}
\ . 
\ee
Since the latter is negative for $\alpha^2 \ll 1$ and $a^2 \ll 1$, 
we expect there exists one horizon whose location is given exactly by
the original (spherically symmetric) solution~\eqref{rha0} for the unperturbed space-time.
This expectation will indeed be confirmed by the study of null geodesics.
\subsection{Trapping surfaces}
Since the $g_{xx}$ component of the metric~\eqref{1} is not well defined at $x^2=1$,
it will be more convenient to work with the usual azimuthal coordinate $\theta$
(with the poles at $\theta=0$ and $\theta=\pi$).
In particular, 
\be
h_a(r,\theta)
\simeq
\left(1-\alpha^{2}\,r\right)
-\frac{a^2\,\alpha^{2}}{4\,r}
\left[3\ln (\alpha^2 \, r)+\cos^{2}\theta\,(1-\,\alpha^2\, r)
\right]
\ .
\label{h3}
\ee
The Lagrangian for a point particle moving on this space-time can be written as
\be
2\,\mathcal{L}
=
-h_a(r,\theta)\,\dot{t}^{2}
+
\frac{r^{2}+a^{2}\,\cos^{2}\theta}{r^{2}+a^{2}}\,
\frac{\dot{r}^{2}}{h_a(r,\theta)}
+\left(r^{2}+a^{2}\,\cos^{2}\theta\right) \dot{\theta}^{2}
+\left(r^{2}+a^{2}\right)\sin^{2}\theta\,\dot{\phi}^{2}
,
\quad
\label{h4}
\ee
where a dot represents the derivative with respect to the parameter $\lambda$
along the trajectories.
Since $t$ and $\phi$ are cyclic variables, one finds the conserved conjugate momenta
\begin{subequations}
\beq
p_{t}
&=&
-h_a(r,\theta)\,\dot{t}
=
-E
\ ,
\label{h4a}
\\
p_{\phi}
&=&
\left(r^{2}+a^{2}\right)\sin^{2}\theta\,\dot{\phi}
=
J
\ ,
\label{h5}
\eeq
\end{subequations}
where $E$ and $J$ are constants.
In particular, one can always set $\dot\phi=J=0$.
\par
In order to establish the existence of radial geodesics with constant $\theta$, we must consider the
dynamical equation for $\theta$, which turns out to be very cumbersome.
By assuming $J=\dot\theta=0$, it takes the following simple form, 
\be
2\left(r^2+a^2\,\cos^2\theta\right)\partial_\theta h_a
+
2\,h_a\,a^2\,\sin\theta\,\cos\theta
=
0
\ .
\label{ddtheta}
\ee
From Eq.~\eqref{h3}, we have
\be
\partial_\theta h_a
\simeq
\frac{a^2\,\alpha^{2}}{2\,r}\,(1-\,\alpha^2\, r)\,\sin\theta\,\cos\theta
\ ,
\ee
and Eq.~\eqref{ddtheta} is clearly satisfied for $\theta=0$ and $\theta=\pi$ (the poles) and $\theta=\pi/2$
(the equatorial plane).
In these cases, we shall show explicitly in sections~\ref{ss:equator} and~\ref{ss:poles} that the trapping surface
occurs precisely where $h_a=0$.
Since $h_a=0$ implies $\partial_\theta h_a=0$, we verify in section~\ref{ss:gen} that the location of trapping surfaces 
is always determined by Eq.~\eqref{h=0}, as we first conjectured.
\subsubsection{Trapping surface on the equatorial plane}
\label{ss:equator}
For determining the radial null geodesics on the equatorial plane, we can set
$\theta=\pi/2$, $\dot\theta=J=0$ and $2\,\mathcal{L}=0$ then reads 
\be
\frac{\dot{r}}{E}
=
\pm\frac{\sqrt{r^{2}+a^{2}}}{r}
\ ,
\label{h7a}
\ee
where the plus (minus) sign is for outgoing (ingoing) geodesics.
From Eqs.~\eqref{h4a} and~\eqref{h7a}, we can write the 4-vectors respectively
tangent to these outgoing and ingoing null trajectories on the equatorial plane as
\begin{subequations}
\beq
\bm{\ell}
&=&
\frac{1}{2} \left[
\bm{\partial _t}
+
h_a(r,\pi/2)\,\frac{\sqrt{r^{2}+a^{2}}}{r} \,\bm{\partial _r}
\right]
\\
\bm{n}
&=&
\frac{1}{h_a(r,\pi/2)} \, \bm{\partial _t}
-\frac{\sqrt{r^{2}+a^{2}}}{r} \,\bm{\partial _r}
\ ,
\label{h9}
\eeq
\end{subequations}
where we multiplied the outgoing null vector by
\be
h_a(r,\pi/2)
\simeq
(1-\alpha^{2}\,r)
-
\frac{3\,a^2\,\alpha^{2}}{4\,r}\,\ln(\alpha^2 \, r)
\ .
\label{h8}
\ee
in order to satisfy the normalisation condition $\ell^{\mu}\,n_{\mu}=-1$.
Since $\ell^{\mu}$ and $n^{\mu}$ are zero for $\mu=2$ and $3$, 
the induced metric $q_{22}=g_{22}$ and $q_{33}=g_{33}$.
Using Eqs.~\eqref{h10}, the outgoing and ingoing expansion scalars are finally given by
\begin{subequations}
\beq
\Theta_{\bm{\ell}}
&=&
\frac{r^{2}+a^{2}/2}{r^{2}\,\sqrt{r^{2}+a^{2}}}\,h_a(r,\pi/2)
=
-\frac{\Theta_{\bm{n}}}{2}\,h_a(r,\pi/2)
\ ,
\qquad
\label{h12}
\\
\nonumber
\\
\Theta_{\bm{n}}
&=&
-\frac{2\,r^{2}+a^{2}}{r^{2}\,\sqrt{r^{2}+a^{2}}}
=
-\frac{2\,\Theta_{\bm{\ell}}}{h_a(r,\pi/2)}
\ .
\label{h13}
\eeq
\end{subequations}
The location of the trapping surfaces is given by $\Theta_{\bm{\ell}}=0$,
which, from Eq.~\eqref{h12} is precisely the same as Eq.~\eqref{h=0} for $x=0$
(that is, $\theta=\pi/2$). 
\subsubsection{Trapping surface at the poles}
\label{ss:poles}
Along the poles, some care must be taken in order to repeat the procedure of the previous section~\ref{ss:equator},
since setting $\theta=0$ (or $\theta=\pi$) from the onset would make some quantities singular.
We then set $\dot\theta=J=0$, expand all relevant quantities for $\theta\simeq 0$ and take the limit $\theta\to 0$
at the end of the calculation.
In this limit, we find the outgoing null 4-vector
\begin{subequations}
\be
\bm{\ell}
=
\frac{1}{2}\left[
\bm{\partial _t}
+
h_a(r,0)
\,\bm{\partial _r} 
\right]
\ ,
\label{h18}
\ee
and the ingoing null 4-vector
\be
\bm{n}
=
\frac{1}{h_a(r,0)}
\bm{\partial _t} 
-\bm{\partial _r}
\ ,
\label{h19}
\ee
\end{subequations}
which satisfy the normalization condition $\ell^{\mu}\,n_{\mu}=-1$ with
\be
h_a(r,0)
\simeq
\left(1-\alpha^{2}\,r\right)
-\frac{a^2\,\alpha^{2}}{4\,r}
\left[3\ln (\alpha^2 \, r)+(1-\,\alpha^2\, r)
\right]
\ .
\ee
The expansion scalars are finally given by
\begin{subequations}
\beq
\Theta_{\bm{\ell}}
=
\frac{r}{r^2+a^2}\,h_a(r,0)
=
-\frac{\Theta_{\bm{n}}}{2}\,h_a(r,0)
\ ,
\eeq
\end{subequations}
and the location of the trapping surface by Eq.~\eqref{h=0} for $x^2=1$ (that is, $\theta=0$ or 
$\theta=\pi$) with the same solution $\rh$ of Eq.~\eqref{rha0}.
\subsubsection{General trapping surface}
\label{ss:gen}
In order to locate the trapping surface for $\theta\not=0$ (or $\theta\not=\pi$) and $\pi/2$, we need to satisfy the
general Eq.~\eqref{ddtheta}.
In analogy with the cases of the poles and equator, we conjecture that Eq.~\eqref{h=0} will be satisfied
and that taking the limit $\dot\theta\to 0$ precisely on the trapping surface is allowed. 
The condition $2\,\mathcal{L}=0$ in this limit reads
\be
\frac{\dot{r}}{E}
=
\pm \sqrt{\frac{r^2+a^2}{r^2+a^2\cos^2\theta_{\rm H}}}
\ ,
\label{gn3}
\ee
where $\theta_{\rm H}$ is the value of the angle $\theta$ on the trapping surface,
where $\dot\theta_{\rm H}=0$.
For $\theta\simeq \theta_{\rm H}$, by using Eqs.~\eqref{h4a} and \eqref{gn3}, we can likewise write the
normalised outgoing null vector as 
\begin{subequations}
\be
\bm{\ell}
\simeq
\frac{1}{2} \left[
\bm{\partial _t}
+
h_a(r,\theta)\,\sqrt{\frac{r^2+a^2}{r^2+a^2\,\cos^2\theta}} \,\bm{\partial _r}
\right] 
\label{gn3a}
\ee
and the normalised ingoing null vector as
\be
\bm{n}
\simeq
\frac{1}{h_a(r,\theta)} \, \bm{\partial _t} 
-\sqrt{\frac{r^2+a^2}{r^2+a^2\,\cos^2\theta}} \, \bm{\partial _r} 
\ ,
\label{gn3b}
\ee
\end{subequations}
which correctly reproduce the previous cases.
Finally, the null expansions in the limit $\dot\theta\to \dot\theta_{\rm H}=0$ are given by
\be
\Theta_{\bm{\ell}}
=
\frac{r\left[2\,r^2+a^2\left(1+\cos^2\theta_{\rm H}\right)\right]}
{2\,\sqrt{(r^2+a^2)(r^2+a^2\,\cos^2\theta_{\rm H})^3}}\,
h_a(r,\theta_{\rm H})
=
-\frac{\Theta_{\bm{n}}}{2}\,h_a(r,\theta_{\rm H})
\label{gn5}
\ ,
\ee
which again reproduces the previous cases.
\par
The condition $\Theta_{\bm \ell}=0$ is therefore fully equivalent to Eq.~\eqref{h=0},
and the location of the trapping surface is given by Eq.~\eqref{rha0} for all values of
$0\le\theta\le\pi$.
Of course, this surface is now a spheroid with deformation parameter $a^2$.
\subsection{Misner-Sharp mass}
In the example considered above, we have found two relevant results:
\begin{description}
\item[a)]
the location of the trapping surface is given by the same value of the radial coordinate
as for the isotropic case.
In particular, we have seen that 
\be
\Theta_{\bm{\ell}}
=
-\frac{\Theta_{\bm{n}}}{2}\,h_a(r,\theta)
\ ,
\ee
for all angles $\theta$ on the trapping surface, and
\item[b)]
$h_a(r,\theta)=0$ where the spherically symmetric $h(r)=h_{a=0}(r,\theta)=0$.
\end{description}
It then follows that the spheroidal horizon is given by
\be
\rh=
2\,m_a(\rh,\theta)
=
2\,m(\rh)
\ ,
\ee
where $m(r)=m_{a=0}(r,\theta)$, and we can conjecture that the relevant mass function
for determining the location of trapping surfaces in (slightly) spheroidal systems is given
by the Misner-Sharp mass computed according to Eq.~\eqref{M} on the reference isotropic
space-time with $a^2=0$~\footnote{This conjecture is further tested in Ref.~\cite{progress}.}.
\section{HQM of deformation parameter} \label{sec-4}
\setcounter{equation}{0}
From the results of the previous section, we expect the HQM for spherically symmetric
sources carries on to the spheroidal case straightforwardly.
For this reason, we shall not report the details here, but just refer to the existing HQM
literature~\cite{fuzzyh,hqm, Giusti:2017byx,gupf,Tevo,cx,otherD,acmo,cc, Giugno:2017xtl}.
Nonetheless, we further need an observable to determine the deformation classically
parametrised by $a$.
\par
For the latter purpose, we introduce a Misner-Sharp mass ``adapted'' to the deformed symmetry
and compare it with the isotropic form~\eqref{M}.
At the classical level, from the flat 3-metric in spheroidal coordinates,
\be
\gamma_{ij}\,\d x^i\,\d x^j
=
\frac{r^{2}+a^{2}\cos{}^{2}\theta}{r^{2}+a^{2}}\,\d r^{2}
+\left(r^{2}+a^{2}\cos^{2}\theta\right) \d\theta^{2}
+\left(r^{2}+a^{2}\right)\sin^{2}\theta\,\d\phi^{2}
\ ,
\ee	
it is easy to see that we can split the general volume measure into two contributions,
and define the adapted Misner-Sharp mass inside spheroids as
\be
\tilde m(r)
=
\int \mu _0 \, \rho
+ a^2 \int \mu _2 \, \rho
\ ,
\label{eq-measure}
\ee 
where $\rho=\rho(r)$ is the energy density,
$\mu _0 (r ; \bar r,  \bar \theta) = \bar r^2\, \sin \bar \theta \, \Theta (r - \bar r)\, \d \bar r\, \d \bar \theta$
is the flat volume measure for spherical domains, and  
$\mu_2 (r ; \bar r,  \bar \theta) = \cos ^2 \bar \theta \,\sin \bar \theta \, \Theta (r - \bar r)\, \d \bar r\, \d \bar \theta$
its deformation. 
\par
At the quantum level, if we wish to tell a spheroidal horizon from a spherical one, from a local perspective~\cite{hqm},
we must analyse the modification to the energy spectrum of the source induced by the spheroidal deformation.
As before, we will only consider small perturbations with respect to a spherical system,  
and replace the adapted Misner-Sharp~\eqref{eq-measure} with the expectation value of the local Hamiltonian
describing the quantum nature of the source,
\be
\hat{H} (r)
=
\hat{H} ^{(0)} (r)
+\frac{a^2}{\Lambda ^2} \, \hat{H}^{(2)} (r)
\ .
\label{eq-Hamiltonian}
\ee
In practice, this quantum prescription implies the following replacements
\begin{subequations} 
\beq
\int \mu _0 \, \rho
\quad
&\overset{\rm HQM}{\longrightarrow}&
\quad
\expec{\hat{H} ^{(0)} (r)}
\\
\int \mu _2 \, \rho
\quad
&\overset{\rm HQM}{\longrightarrow}&
\quad
\Lambda^{-2}\,\expec{\hat{H}^{(2)}(r)}
\ ,
\eeq
\end{subequations} 
where $\Lambda\simeq r_0$ is such that $\epsilon \equiv a / \Lambda \ll 1$.
One can then try and infer the structure of the spectrum
$\sigma (\hat{H} (r)) = \left\{ E_\alpha (r) \, | \, \alpha \in \mathcal{I} \right\}$,
with $\mathcal{I}$ a discrete set of labels (due to the localised nature of the source),
from the spectrum of the corresponding spherical system
$\sigma (\hat{H}^{(0)} (r)) = \left\{ E ^{(0)} _\alpha (r) \, | \, \alpha \in \mathcal{I} \right\}$
using standard perturbation theory.
We expand the solution of the eigenvalue problem for $\hat{H} (r)$ as a Taylor series
in $\epsilon$, namely
\begin{subequations} 
\be 
E_\alpha (r)
=
E ^{(0)} _\alpha (r) + \epsilon^2\, E ^{(2)} _\alpha (r) + \ldots 
\ee
and, omitting the radial dependence for the sake of brevity, we write the eigenvectors as
\be 
\ket{E_\alpha}
=
\ket{E ^{(0)} _\alpha} + \epsilon^2 \, \ket{E ^{(2)} _\alpha} + \ldots
\ .
\ee
\end{subequations} 
\par 
At order $\epsilon^2$, the perturbative solution is given by
\begin{subequations} 
\be 
E_\alpha
\simeq
E ^{(0)}_\alpha
+ \epsilon^2\, \expec{E ^{(0)} _\alpha \ | \ \hat{H} ^{(2)} (r) \ | \ E ^{(0)} _\alpha} 
\label{E}
\ee
and
\be 
\ket{E_\alpha}
\simeq
\ket{E ^{(0)} _\alpha}
+ 
\epsilon^2 \,
\sum_{\beta \neq \alpha} 
\frac{\expec{E ^{(0)} _\beta \ | \ \hat{H} ^{(2)} (r) \ | \ E ^{(0)} _\alpha}}{E ^{(0)} _\alpha - E ^{(0)} _\beta}
\,\ket{E ^{(0)} _\beta}
\ .
\label{ket} 
\ee
\end{subequations}
The above expressions allow us to introduce a characterisation of the deformation parameter $a^2$
in a purely quantum framework.
\par
In particular, Eq.~\eqref{E} implies
\be 
\frac{a^2}{\Lambda ^2}
\simeq
\frac{E_\alpha - E ^{(0)} _\alpha}{\expec{E ^{(0)} _\alpha \ | \ \hat{H} ^{(2)} (r) \ | \ E ^{(0)} _\alpha}}
\ .
\ee
Moreover, given a quantum state for the source, we can express it either using the deformed spectrum 
$\sigma (\hat{H} (r))$,
\be
\ket{\psi}
=
\sum_\alpha C_\alpha (r) \,\ket{E_\alpha}
\ ,
\ee
or the isotropic spectrum $\sigma (\hat{H}^{(0)} (r))$, 
\be
\ket{\psi}
=
\sum_\alpha C_\alpha^{(0)} (r) \,\ket{E_\alpha^{(0)}}
\ ,
\ee
and Eq.~\eqref{ket} yields
\be
C_\alpha (r) - C_\alpha^{(0)} (r)
\simeq
\frac{a^2}{\Lambda ^2}\,
\sum_{\beta \neq \alpha} 
\frac{\expec{E ^{(0)} _\beta \ | \ \hat{H} ^{(2)} (r) \ | \ E ^{(0)} _\alpha}}{E ^{(0)} _\alpha - E ^{(0)} _\beta}
\,C^{(0)} _\beta (r)
\ .
\ee
\par
We now recall that the spectrum of the gravitational radius operator is related to the source through the
Hamiltonian constraint~\cite{fuzzyh,hqm}
\be
\left({\hat R}_{\rm H} (r) - \frac{2 \, \lp}{\mpl} \, \hat{H} (r) \right)
\ket{\Psi}
=
0
\, ,
\ee
where $\ket{\Psi} = \sum _\alpha \mathcal{C}_{\alpha, \beta} \ket{E _\alpha} \ket{R_{{\rm H}, \beta}}$,
and one therefore finds $\mathcal{C} _{\alpha, \beta} = C _\alpha (r) \, \delta _{\alpha, \beta}$.
One can then infer how the spheroidal deformation affects the spectrum of the
gravitational radius and the form of the horizon wave function.
Indeed, by means of the Hamiltonian constraint one can immediately see that 
\be 
R_{{\rm H}, \alpha}
\simeq
R ^{(0)} _{{\rm H}, \alpha}
+ \epsilon^2\,
\expec{R ^{(0)} _{{\rm H}, \alpha} \, | \, {\hat R} ^{(2)} _{\rm H}  (r) \, | \, R ^{(0)} _{{\rm H}, \alpha}}
\, ,
\label{E}
\ee
with ${\hat R} ^{(2)} _{\rm H}  (r) \equiv 2 \, \lp \, \hat{H} ^{(2)} (r) /\mpl$.
Furthermore, recalling that the state of the geometry in this language can be obtained by tracing away
the contribution of the source from the general state $\ket{\Psi}$~\cite{fuzzyh,hqm},
one obtains
\be 
\ket{\psi _{\rm H}}
=
\sum _\alpha C_{\alpha} (r) \, \ket{R_{{\rm H}, \alpha}}
\ ,
\ee
and the horizon wave function is finally given by
\beq 
\psi _{\rm H} (R_{{\rm H}, \alpha})
&\!\!=\!\!&
\expec{R ^{(0)} _{{\rm H}, \alpha} \ | \ \psi _{\rm H}}
=
C_\alpha (r)
\nonumber
\\
&\!\!\simeq\!\!&
C ^{(0)} _\alpha (r) + \frac{a^2}{\Lambda ^2}\,
\sum_{\beta \neq \alpha} 
\frac{\expec{R ^{(0)} _{{\rm H}, \beta} \, | \, \hat{H} ^{(2)} (r) \, | \, R ^{(0)} _{{\rm H}, \alpha}}}
{R ^{(0)} _{{\rm H}, \alpha} - R ^{(0)} _{{\rm H}, \beta}}
\,C^{(0)} _\beta (r)
\ .
\label{hwfa}
\eeq
It will be interesting to apply the above treatment to specific cases, and also include rotation~\cite{rotating},
which we plan to consider in future publications.
\section{Conclusions}
After reviewing some general aspects of the theory of trapping surfaces is General Relativity,
with particular regard to static and spherically symmetric systems, we have described a general
procedure for studying the location of these surfaces in static and slightly spheroidal
space-times (that is, for small deformation parameter $a^2$ in Eq.~\eqref{1}).
\par
More specifically, we started in Sec.~\ref{Sec:spheroidal} with some general remarks on the
geometry and energy density of a generic static source with slightly spheroidal profile.
In Sec.~\ref{sec:3}, we then studied a slightly spheroidal de~Sitter space generated by
an energy density $\rho \propto 1/r$, for which we provided a precise characterisation
of the structure of the trapping surfaces.
The main result of this analysis is that the location of a trapping surface appears to be
(analytically) the same function of the radial coordinate $r$ for both the isotropic and slightly
spheroidal systems (where $r$ is constant on surfaces of the respective symmetries).
Moreover, the discussion suggests that the appropriate generalisation of the Misner-Sharp mass,
for determining the location of these hypersurfaces for small $a^2$, is the the usual Misner-Sharp
mass computed for the isotropic space (that is, for $a^2 = 0$).
Clearly, this is a conjecture we have verified for a particular slightly spheroidal system,
and it would be surely worth understanding how to extend it to more general systems 
(for instance, with large deformation parameter $a^2$). 
This is left for future investigations (see also Ref.~\cite{progress}).
\par
Finally, in Sec.~\ref{sec-4} we have qualitatively discussed the deformations induced
on the spectrum of the isotropic source by these small spheroidal deformations within the
framework of quantum perturbation theory.
This standard procedure allowed us to describe the effects of the deformation parameter $a$
on the spectrum of the operator associated with the gravitational radius of the system (see Eq.~\eqref{E}),
as well as on the local notion of the horizon wave function~\eqref{hwfa}, which represents the most
important outcome of the HQM formalism.
\section*{Acknowledgments}
R.C.~and A.G.~are partially supported by the INFN grant FLAG and
their work has been carried out in the framework of GNFM and INdAM
and the COST action {\em Cantata\/}. 
R.R.~is supported by a IRSIP grant of the HEC.
\appendix

\begin{thebibliography}{99}
%
%
%
\bibitem{Wald} 
 R.~M.~Wald,
 ``General Relativity,''
(University of Chicago Press, USA, 1984)
%
\bibitem{HE} 
S.~W.~Hawking and G.~F.~R.~Ellis,
``The Large Scale Structure of Space-Time,''
(Cambridge University Press, Cambridge, UK, 1973)
%
\bibitem{schwarz}
K.~Schwarzschild,
Sitzungsber.\ Preuss.\ Akad.\ Wiss.\ Berlin (Math.\ Phys.\ ) {\bf 1916} (1916) 189
 [physics/9905030].
%
\bibitem{ashtekar}
A.~Ashtekar and B.~Krishnan,
Living Rev.\ Rel.\  {\bf 7} (2004) 10
[gr-qc/0407042].
%
\bibitem{hawking}
S.~W.~Hawking,
Commun.\ Math.\ Phys.\  {\bf 43} (1975) 199
Erratum: [Commun.\ Math.\ Phys.\  {\bf 46} (1976) 206].
%
\bibitem{fuzzyh}
R.~Casadio,
``Localised particles and fuzzy horizons: A tool for probing Quantum Black Holes,''
arXiv:1305.3195 [gr-qc];
``What is the Schwarzschild radius of a quantum mechanical particle?,''
arXiv:1310.5452 [gr-qc].
%
\bibitem{hqm}
R.~Casadio, A.~Giugno and A.~Giusti,
Gen.\ Rel.\ Grav.\  {\bf 49} (2017) 32
[arXiv:1605.06617 [gr-qc]].
\bibitem{Giusti:2017byx} 
 A.~Giusti,
 J.\ Phys.\ Conf.\ Ser.\  {\bf 942} (2017) 012013
 [arXiv:1709.10348 [gr-qc]].
%
\bibitem{gupf} 
R.~Casadio and F.~Scardigli,
Eur.\ Phys.\ J.\ C {\bf 74} (2014) 2685
[arXiv:1306.5298 [gr-qc]].
%
\bibitem{Tevo}
R.~Casadio,
Eur.\ Phys.\ J.\ C {\bf 75} (2015) 160
[arXiv:1411.5848 [gr-qc]].
%
\bibitem{cx}
X.~Calmet and R.~Casadio,
Eur.\ Phys.\ J.\ C {\bf 75} (2015) 445
[arXiv:1509.02055 [hep-th]].
%
\bibitem{otherD}
R.~Casadio, R.~T.~Cavalcanti, A.~Giugno and J.~Mureika,
Phys.\ Lett.\ B {\bf 760} (2016) 36
[arXiv:1509.09317 [gr-qc]].
%
\bibitem{acmo}
R.~Casadio, O.~Micu and F.~Scardigli,
Phys.\ Lett.\ B {\bf 732} (2014) 105
[arXiv:1311.5698 [hep-th]].
%
\bibitem{cc}
R.~Casadio, O.~Micu and D.~Stojkovic,
Phys.\ Lett.\ B {\bf 747} (2015) 68
[arXiv:1503.02858 [gr-qc]];
JHEP {\bf 1505} (2015) 096
[arXiv:1503.01888 [gr-qc]].
%
\bibitem{Giugno:2017xtl} 
A.~Giugno, A.~Giusti and A.~Helou,
``Horizon quantum fuzziness for non-singular black holes,''
 arXiv:1711.06209 [gr-qc].
%
\bibitem{rotating}
R.~Casadio, A.~Giugno, A.~Giusti and O.~Micu,
Eur.\ Phys.\ J.\ C {\bf 77} (2017) 322
[arXiv:1701.05778 [gr-qc]].

%
\bibitem{stephani}
H.~Stephani,
``Relativity: An introduction to special and general relativity,''
(Cambridge University Press, Cambridge, UK, 2004)
%
\bibitem{progress}
R.~Rahim, A.~Giusti, R.~Casadio,
``On trapping surfaces in spheroidal space-times,''
in preparation.
%
\end{thebibliography}
\end{document}